# An Energy-Efficient RFET-Based Stochastic Computing Neural Network Accelerator

Sheng Lu, *Student Member*, *IEEE*, Qianhou Qu, *Student Member*, *IEEE*, Sungyong Jung, *Member*, *IEEE*, Qilian Liang, *Fellow*, *IEEE*, and Chenyun Pan, *Senior Member*, *IEEE*

*Abstract*—Stochastic computing (SC) offers significant reductions in hardware complexity for traditional convolutional neural networks (CNNs). However, despite its advantages, stochastic computing neural networks (SCNNs) often suffer from high resource consumption due to components such as stochastic number generators (SNGs) and accumulative parallel counters (APCs), which limit overall performance. This paper proposes a novel SCNN accelerator based on reconfigurable field-effect transistors (RFETs). The inherent reconfigurability at the device level enables the design of highly efficient and compact SNGs, APCs, and other related essential components. To assess their system-level impact, a representative existing SCNN architecture is adopted as an evaluation framework. Based on accessible open-source standard cell libraries, experimental results demonstrate that the proposed RFET-based SCNN accelerator achieves significant reductions in area, latency, and energy consumption compared to its FinFET-based counterpart at the same technology node.

## I. Introduction

With the advancement of neural network technologies, various alternative computing paradigms have been proposed to address the growing demand for efficient hardware implementations [1, 2]. Among these approaches, the stochastic computing neural network (SCNN) emerges as a promising solution due to its fundamental differences from conventional binary computing [3-6]. SCNN enables low-cost hardware implementations, offering high power efficiency and inherent fault tolerance. It significantly reduces the hardware complexity of fundamental arithmetic units, such as adders and multipliers, which typically occupy a substantial portion of the system's overall computing resources.

SCNNs typically require a large number of stochastic number generators (SNGs) [7, 8], each consisting of a random number source (RNS) and a probability conversion circuit (PCC) to convert binary-encoded values into stochastic bitstreams. SNGs can occupy up to 90% of the total system area, as reported in previous studies [9-11]. To mitigate this overhead, various optimization techniques have been proposed. For the RNS component, one of the most effective methods is RNS sharing, which reduces area and cost by distributing the output bitstreams of linear feedback shift registers (LFSRs) across multiple SNGs via signal shuffling. In contrast, optimizing the PCC is more challenging. The traditional comparator-based (CMP) PCC, while simple, incurs high area overhead. To address this, alternative designs such as the weighted binary generator (WBG) and the multiplexer chain (MUX-chain) are proposed [8]. Although the MUX-chain offers the best improved hardware efficiency, the PCC still dominates both the area and energy consumption.

Besides SNGs, another key component in stochastic computing is the accumulative parallel counter (APC), which is widely used in the stochastic domain. Unlike the adders employed in conventional digital neural networks, the APC is used to accumulate multiplication results produced by stochastic multipliers while converting stochastic bitstreams into binary values, thereby serving as a bridge between the stochastic and binary domains. An APC typically consists of full adders (FAs) and half adders (HAs) arranged in a carry-propagation–like topology. Although this structure provides high accuracy and a straightforward implementation, it usually incurs a relatively large area overhead. To address this issue, the approximate accumulative parallel counter (AxPC) is proposed [12]. In AxPC designs, simple logic gates are used in the first stage to perform parity (odd–even) checking instead of exact addition, which reduces hardware complexity at the cost of reduced computational accuracy.

In addition to circuit-level optimization, beyond-CMOS technologies exhibit significant potential in addressing the challenge [13-17]. Among emerging device technologies, the reconfigurable field-effect transistor (RFET) is particularly well-suited for this application. Owing to its inherent device-level ambipolarity—enabling dynamic switching between p-type and n-type operations [18-21], RFETs are well-suited for implementing various digital logic components, including full adders (FAs) [22-24], look-up tables (LUTs) [25], multiplexers [26], and so on. Existing studies have explored the design of key neural network components based on RFETs, such as approximate compressors and multipliers [16, 27]. However,

This work is funded by the Advanced Scientific Computing Research (ASCR) program of the Department of Energy (DOE) through award DE-SC0022881, and National Science Foundation (NSF) under grant CCF-2219753.
S. Lu, Q. Qu, Q. Liang, and C. Pan are with the Department of Electrical Engineering, The University of Texas at Arlington, Arlington, TX 76010, USA. (email: sxl2408@mavs.uta.edu)
S. Jung is with the Department of Electrical Engineering and Computer Science, South Dakota State University, Brookings, SD 57007, USA.

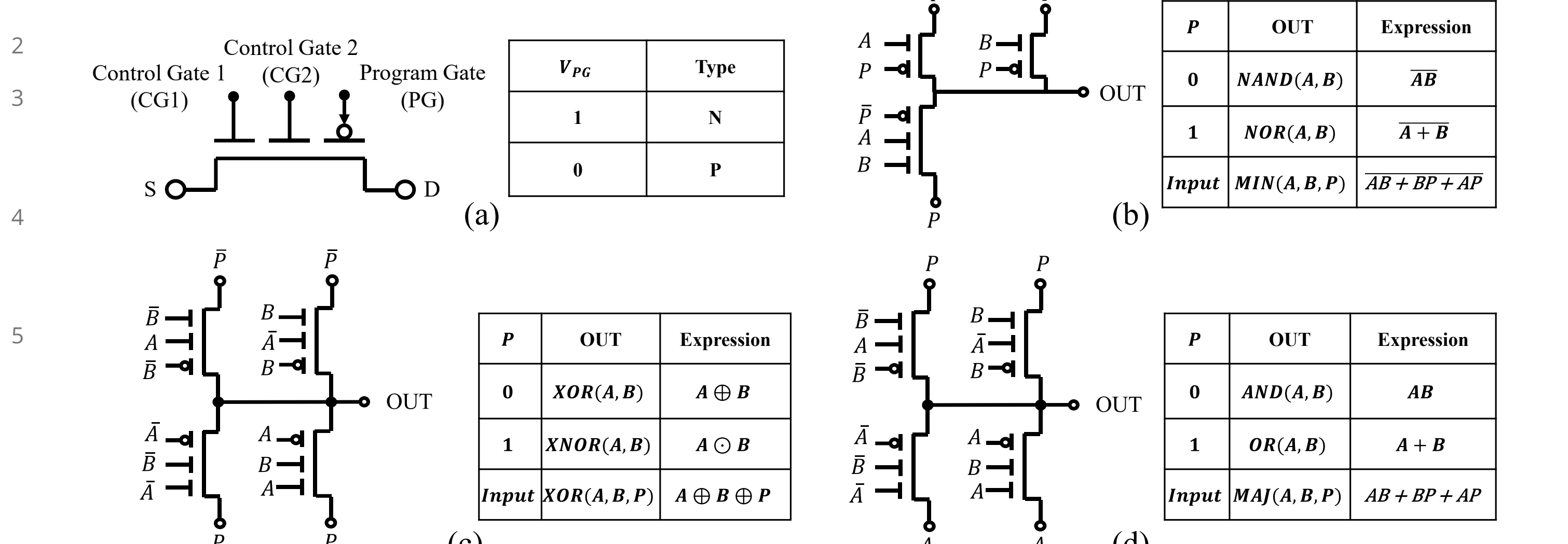

| $V_{PG}$ | Type |
|---|---|
| 1 | N |
| 0 | P |

| P | OUT | Expression |
|---|---|---|
| 0 | $NAND(A, B)$ | $\overline{AB}$ |
| 1 | $NOR(A, B)$ | $\overline{A+B}$ |
| Input | $MIN(A, B, P)$ | $\overline{AB+BP+AP}$ |

| P | OUT | Expression |
|---|---|---|
| 0 | $XOR(A, B)$ | $A \oplus B$ |
| 1 | $XNOR(A, B)$ | $A \odot B$ |
| Input | $XOR(A, B, P)$ | $A \oplus B \oplus P$ |

| P | OUT | Expression |
|---|---|---|
| 0 | $AND(A, B)$ | $AB$ |
| 1 | $OR(A, B)$ | $A+B$ |
| Input | $MAJ(A, B, P)$ | $AB+BP+AP$ |

**Fig. 1.** (a) Symbol of a symmetrical RFET, where PG acts as polarity-program gate and CG1/CG2 serve as the switching gates. Applying different voltages to PG configures the device for p-type or n-type operation. (b–d) Examples of multi-gate RFET structures illustrating how additional independent gates enable higher-order logic such as MIN3 (minority), XOR3 (odd parity) and MAJ3 (majority) [16, 22]. The tables show definitions of these logic gates, where $A \oplus B$ denotes XOR(A, B), $A \odot B$ denotes XNOR(A, B). The pin with the small circle is the program gate PG, while the others are control pins CG1/CG2.

these works primarily focus on optimizing traditional digital neural networks and circuits rather than SCNNs. For the SNG, we demonstrate that an RFET NAND–NOR chain–based PCC achieves significant improvements in area, delay, and energy efficiency while maintaining accuracy. For the APC part, two AxPC architectures based on RFET high-efficiency FAs and compressors are proposed. The first design employs RFET 3-input majority gates as the input layer to further reduce area, delay, and energy at the cost of a slight loss in accuracy, whereas the second design is based on an RFET 4:2 compressor input layer, which significantly improves accuracy while preserving the performance advantage.

In this work, an RFET-based SCNN accelerator is proposed, demonstrating notable performance improvements over FinFET-based counterparts at the same technology node. Two primary functional units are investigated: the RFET-based PCC and the RFET-based AxPC. The key contributions of this paper are outlined as follows:

- We design a novel PCC structure based on RFET reconfigurable NAND-NOR gates with theoretical analysis, achieving reduced area, delay, and energy consumption compared to its FinFET-based counterpart.
- We propose two RFET-based AxPC architectures based on high-efficiency RFET logic units, which can be applied in different scenarios and offer flexibility in accuracy and hardware efficiency trade-offs.
- We use an established SCNN system architecture as an evaluation framework and perform a fair system-level comparison between RFET-based and FinFET-based implementations.

## II. Background

### A. RFET Basic

RFETs are characterized by multiple gate terminals, including one program gate (PG) that controls the polarity of the transistor, and two control gates (CGs) that regulate its switching behavior [18, 21]. **Fig. 1**(a) illustrates the symbol of a tri-gate RFET and its ambipolar operating principle. In this device, PG functions as a polarity-control gate, while the CG1 and CG2 serve as the switching gates. The RFET turns ON only when all gate terminals are biased at compatible voltage levels. A key property of RFETs is that their ON-state resistance is dominated by the source-side Schottky barrier, so adding multiple gate terminals has only a marginal impact on the drive current [28, 29]. Moreover, RFETs exhibit extremely low leakage currents [30], making them highly attractive for energy-efficient electronic systems [21]. According to recent fabrication studies, RFETs are compatible with standard CMOS process technologies, facilitating seamless integration into current nanoscale CMOS manufacturing processes without substantial changes to materials or processing steps [31, 32]. Nonetheless, limitations such as relatively lower drive current in the ON state and a comparatively larger device footprint may hinder RFETs from fully replacing conventional CMOS transistors in all applications.

A direct benefit of this ambipolarity is the ability to design multi-functional logic gates. For example, as shown in **Fig. 1**(b), a three-transistor gate can switch between NAND and NOR logic functions based on the applied P-node voltage. When utilizing the program gate P as the third input, RFET can build compact reconfigurable three-input logic gates, such as MIN3 (minority), XOR3 (odd-parity) and MAJ3 (majority), as shown in **Fig. 1**(b), (c) and (d). This device-level reconfigurability enables more efficient and compact circuit designs, making RFETs particularly well-suited for area- and power-constrained applications.

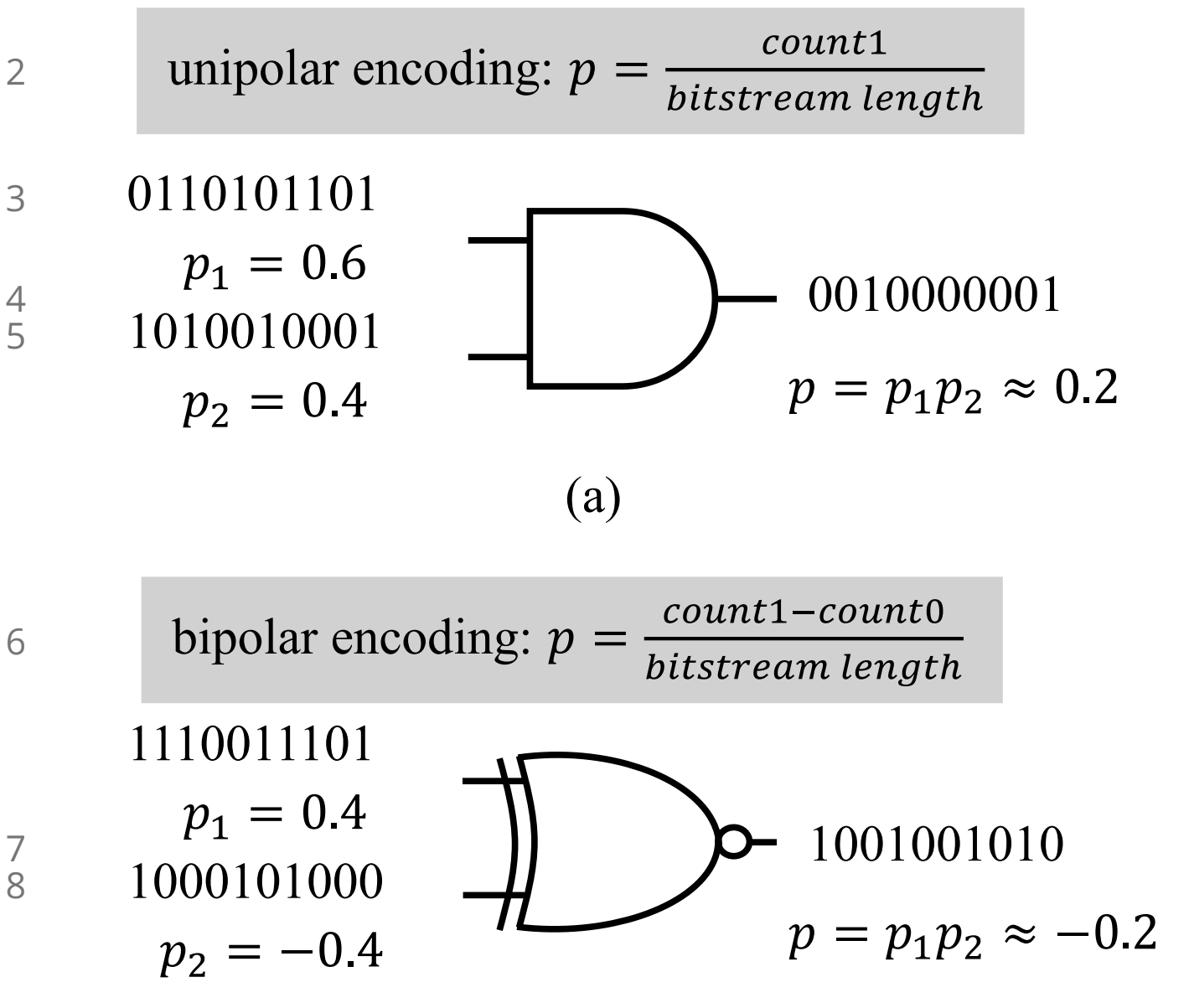

**Fig. 2.** Multiplication based on stochastic computing [4]. (a) Unipolar encoding with AND gate, and (b) bipolar encoding with XNOR gate.

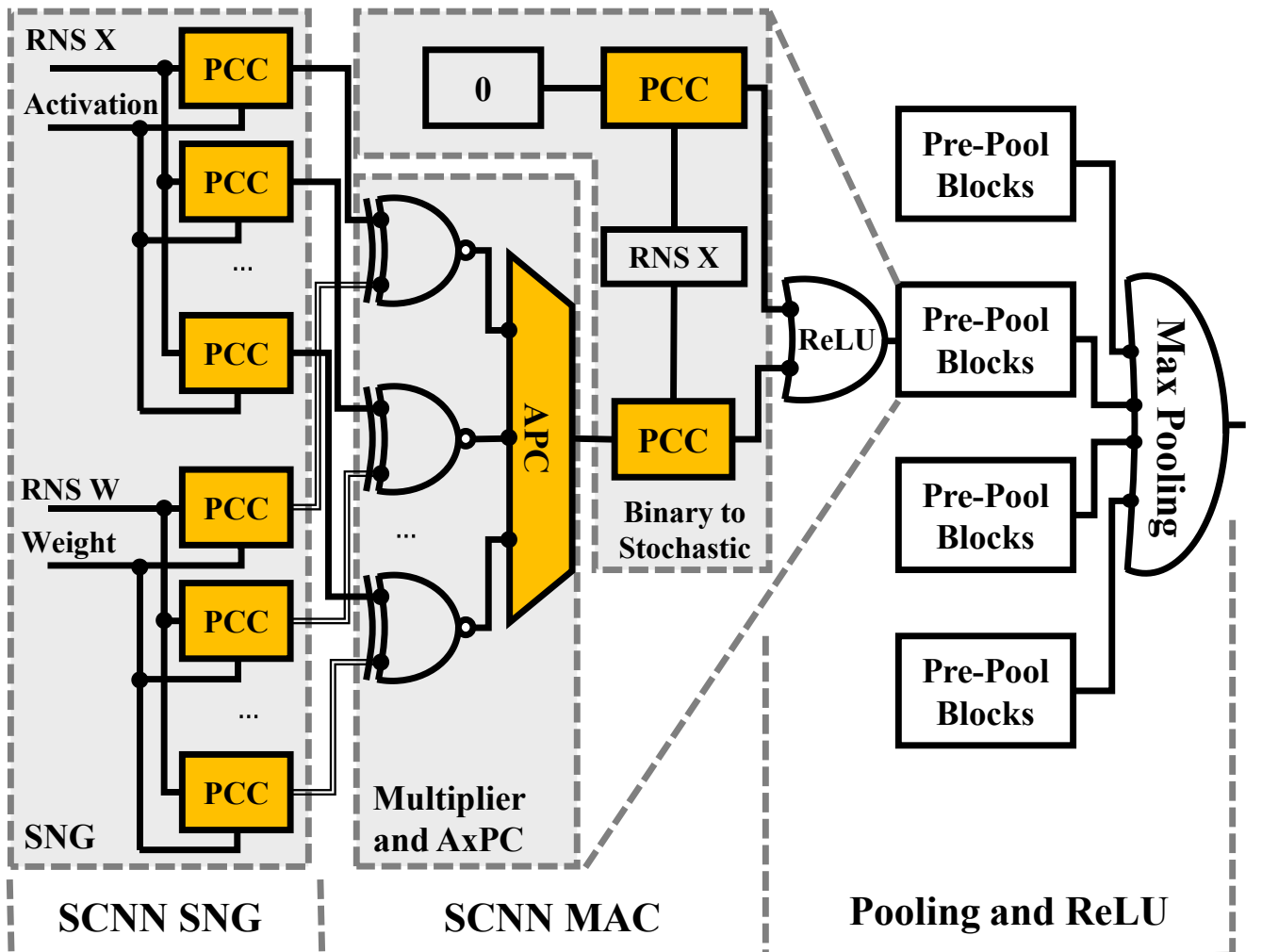

**Fig. 3.** Stochastic neuron proposed by Frasser et al. with fully correlated inputs to both ReLU and MP components to reduce hardware area [36]. This work mainly focuses on the RFET design of the logic blocks highlighted in orange. While all logic elements are RFET-based, these modules are the most critical in the SCNN and best demonstrate the performance benefits of RFET.

### *B. Stochastic Computing Basic*

Unlike conventional binary computation, stochastic computing performs arithmetic operations using sequential bitstreams, where the number of '1's and '0's—corresponding to high and low voltage levels—encodes numerical values. Two primary encoding schemes are commonly employed: unipolar encoding, which represents values in the range [0, 1], and bipolar encoding, which represents values in the range [−1, 1] [3]. The fundamental concepts of both encoding schemes are illustrated in **Fig. 2**. For instance, to represent the value 0.4 in unipolar encoding, a bitstream containing 40% randomly distributed '1's suffices.

Due to the nature of stochastic representation, the corresponding arithmetic circuits in SC differ significantly from those used in traditional binary logic. As shown in **Fig. 2**, multiplication can be implemented using a simple AND gate in the unipolar domain or an XNOR gate in the bipolar domain. Similarly, addition can be realized using lightweight logic elements, such as OR gates or MUXes, enabling highly area- and energy-efficient circuit designs [4]. Furthermore, by appropriately combining or utilizing certain logic gates, more complex computations—such as proportional scaling and approximation of unary linear functions—can be achieved [33], thereby offering greater potential for implementing neural network-related operations. Although this work focuses on unipolar stochastic representation, RFET logic is also promising for bipolar stochastic computing because XNOR-based multiplication can be implemented efficiently using RFET reconfigurable logic.

### *C. Stochastic Computing Neuron Structure*

Although numerous convolutional neural network (CNN) architectures exist, the fundamental building block of these models is the neuron. A CNN neuron is a computational unit characterized by local receptive fields, shared weights, and a nonlinear activation function, enabling the extraction of local features from input images or data [34]. At the hardware level, a neuron may be implemented using multipliers and adders to perform convolutional or multiplicative operations [35]. Activation functions, such as Rectified Linear Unit (ReLU) or Sigmoid, can be realized through corresponding dedicated logic circuits. Furthermore, pooling operations, commonly employed in CNNs, serve as downsampling mechanisms that reduce the spatial dimensions of feature maps while retaining essential information.

The basic neuron structure in SCNNs follows a similar architecture, as shown in **Fig. 3**. This design, proposed by Frasser et al., presents an SC neuron that leverages input correlation to reduce hardware area costs [36]. The neuron employs two RNSs in total. One RNS is used to encode the activation values and to generate correlated input streams for the ReLU and Max Pooling (MP) operations. The other RNS is used to encode the weights. In contrast to traditional digital neurons, which typically employ adder tree structures to accumulate the outputs from multiple multipliers, SC neurons utilize an APC to perform the summation. The APC operates on the input bits at each clock cycle and outputs the count of logic '1's in binary format. Following this, a binary-to-stochastic (B2S) converter is used to transform the binary sum back into stochastic format, enabling continued processing within the SC domain. As previously discussed, in stochastic computing, an OR gate can function as an adder when the input bitstreams are statistically independent. However, when the bitstreams are fully correlated—such as when generated from the same RNS—the OR gate tends to behave more like a maximum operator. This behavior can be exploited to implement ReLU and MP functions efficiently.

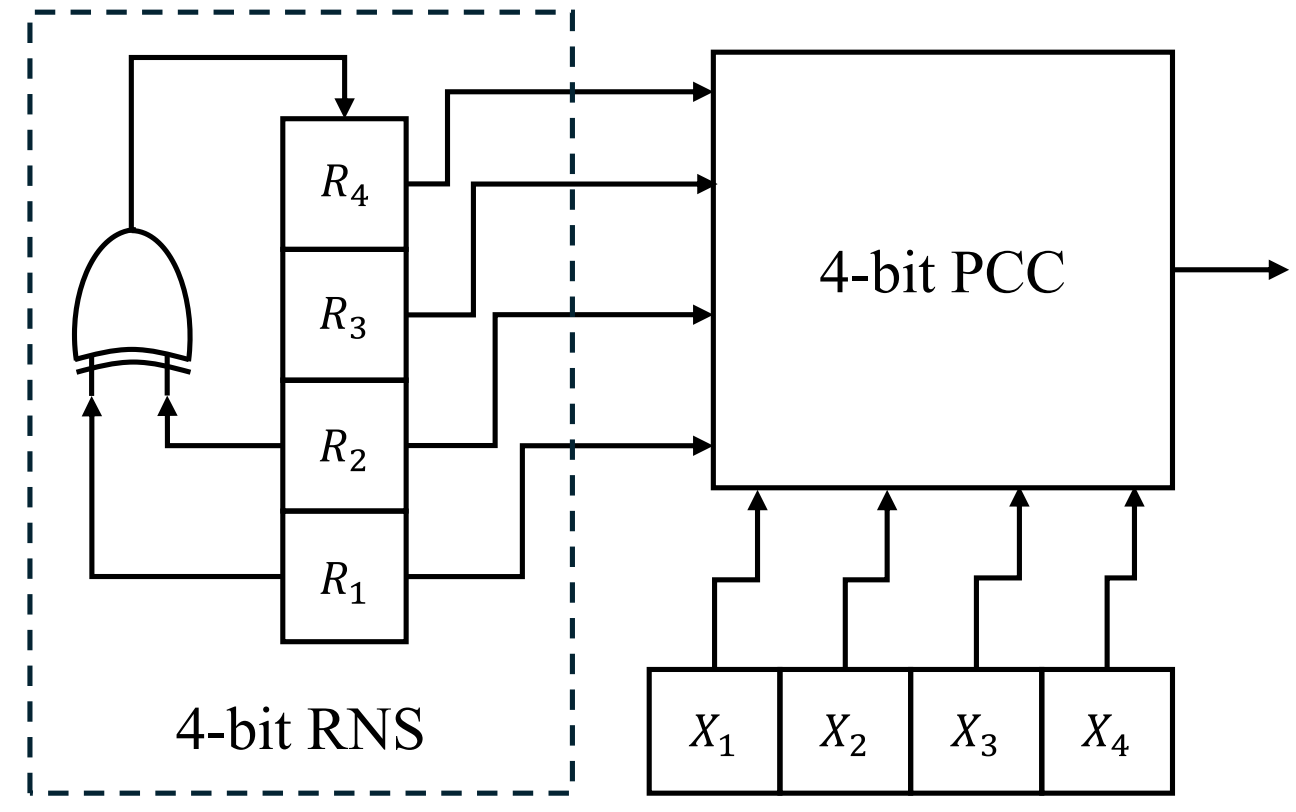

**Fig. 4.** Example circuit schematic of a 4-bit SNG [11].

The neuron structure shown in **Fig. 3** not only reduces the hardware area required for addition and multiplication but also simplifies the implementation of activation and pooling functions, which typically demand significantly more area in conventional digital neuron designs. Due to these advantages, this structure is adopted as the fundamental building block of the proposed SCNN accelerator.

*D. Stochastic Number Generator*

The SNG consists of two main components: an RNS and a PCC. The most used RNS is the LFSR, which generates pseudo-random sequences by shifting bits through a register and applying a linear function—typically an XOR—on selected bits to determine the new input bit. An LFSR can produce a maximum-length pseudo-random sequence of $2^n - 1$ distinct states for an n-bit register when its feedback polynomial is primitive, ensuring that it cycles through all possible non-zero states before repeating. As illustrated in **Fig. 4**, which shows an example of a 4-bit SNG, the left side consists of a 4-bit LFSR, while the right side features a 4-bit PCC [11]. The PCC generates one stochastic bit per clock cycle based on the binary value from the LFSR and the given input.

**Fig. 5** presents two typical implementations of the PCC [10, 11]. **Fig. 5**(a) shows the traditional comparator design, which compares the input binary number with the random number generated by the RNS cycle by cycle. If the input number X is greater than the random number R, the output is logic ‘1’; otherwise, the output is logic ‘0’. **Fig. 5**(b) illustrates the MUX-chain-based PCC. Unlike the comparator-based design, the MUX-chain PCC does not perform a direct comparison between the input and the random number. Instead, it operates probabilistically by selecting signals through a chain of multiplexers. Assuming that the random bits are independent, the output probability depends on the configuration of the MUX-chain and the input value. The output probability can be expressed as [10]:

$$P = \frac{\sum_{i=0}^{N-1} X_i 2^i}{2^N} \quad (1)$$

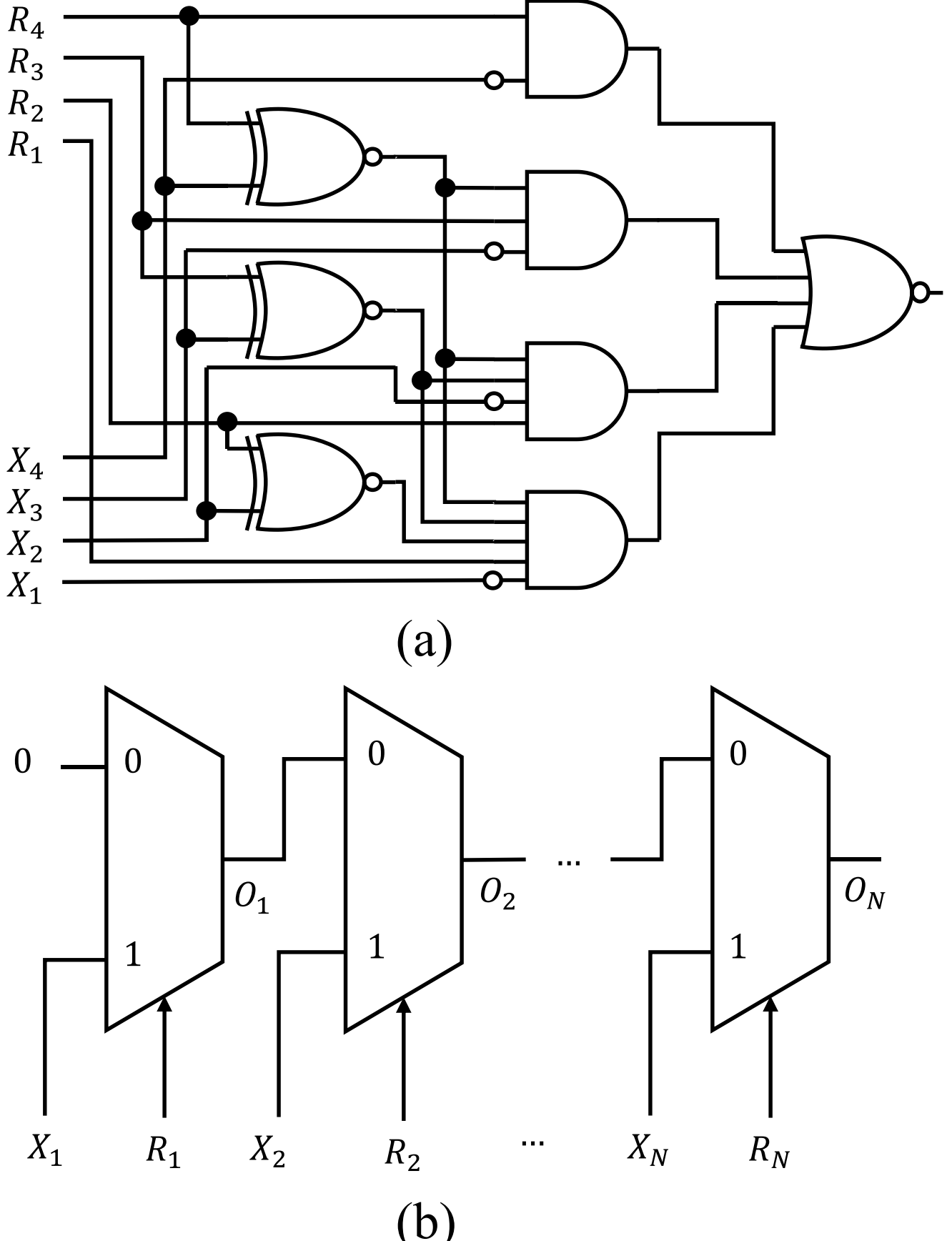

**Fig. 5.** Commonly used PCCs. (a) CMP-based 4-bit PCC [11], and (b) MUX-chain-based N-bit PCC [10].

Based on this expression, the MUX-chain PCC effectively transforms the input binary number into a probability value between 0 and 1, depending on the value of the input. With a sufficiently long bitstream, the output closely approximates the corresponding real-valued representation of the input.

A key advantage of the MUX-chain-based PCC is its significant reduction in hardware area. Compared to the comparator-based design, it can achieve area savings of up to 43% and 58% in 4-bit and 8-bit configurations [11], respectively. Thanks to its efficient and compact nature, it has been adopted for certain chip tape-out works [37]. However, this architecture involves a trade-off between computation precision and critical-path delay. As the required precision increases, i.e., when a higher-bit PCC is adopted, the multiplexer chain becomes longer, resulting in an increased critical-path delay.

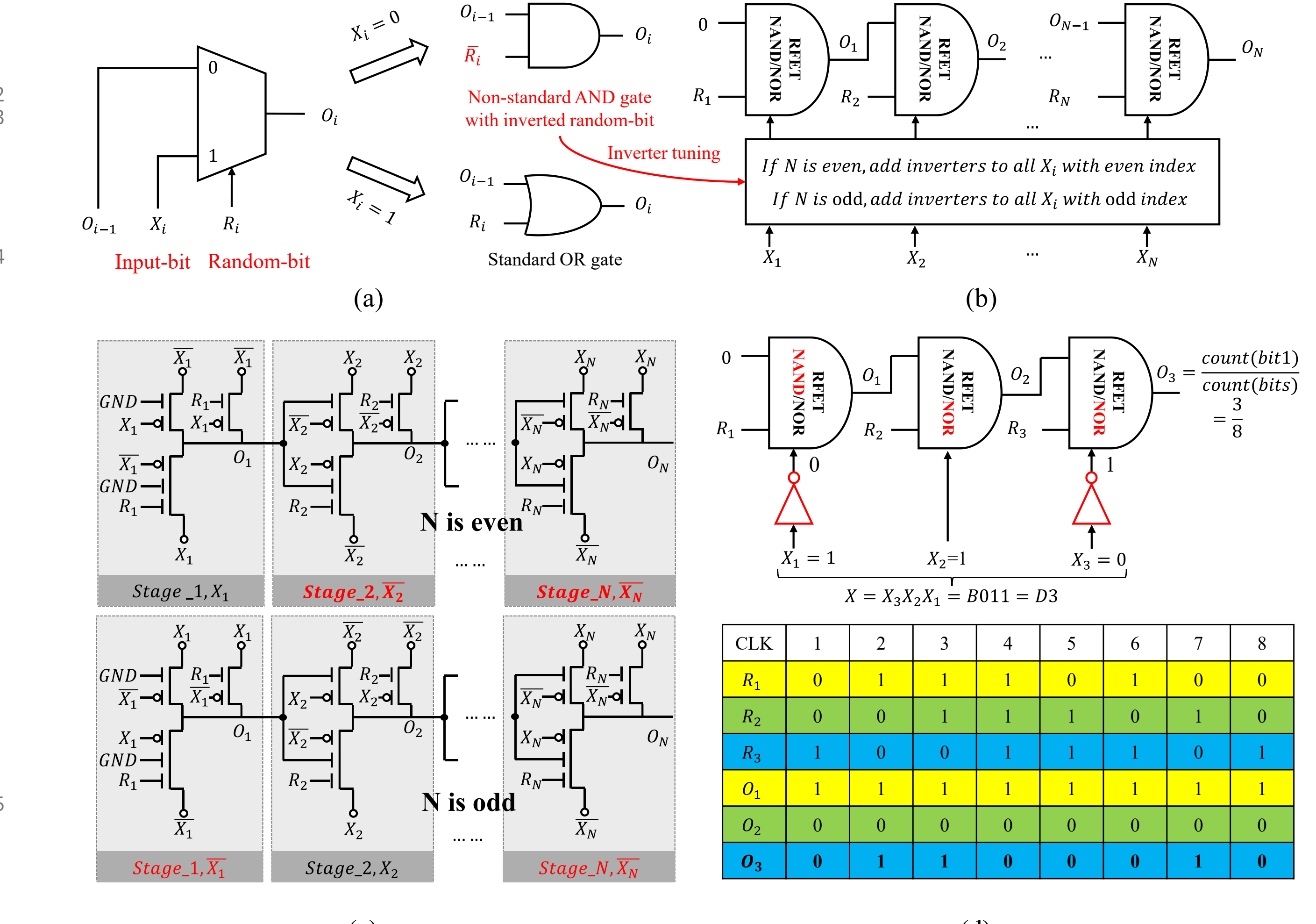

| CLK | 1 | 2 | 3 | 4 | 5 | 6 | 7 | 8 |
| --- | --- | --- | --- | --- | --- | --- | --- | --- |
| $R_1$ | 0 | 1 | 1 | 1 | 0 | 1 | 0 | 0 |
| $R_2$ | 0 | 0 | 1 | 1 | 1 | 0 | 1 | 0 |
| $R_3$ | 1 | 0 | 0 | 1 | 1 | 1 | 0 | 1 |
| $O_1$ | 1 | 1 | 1 | 1 | 1 | 1 | 1 | 1 |
| $O_2$ | 0 | 0 | 0 | 0 | 0 | 0 | 0 | 0 |
| $O_3$ | 0 | 1 | 1 | 0 | 0 | 0 | 1 | 0 |

**Fig. 6.** (a) Equivalent logic representation of the 2:1 multiplexer (MUX21) using AND/OR gates. (b) RFET-based NAND–NOR PCC with tuned inverters. (c) Detailed circuit implementation of the NAND–NOR PCC for even and odd numbers of inputs. (d) Example of a 3-bit PCC over eight clock cycles with the activation input vector 011. The random bits are generated by the corresponding 3-bit LFSR.

## III. Proposed RFET-based Stochastic Computing

### *A. RFET-based Probability Conversion Circuit*

As shown in **Fig. 5**(b), an $i$-stage MUX21 logic gate in the MUX-chain has the following logic expression:

$$MUX(O_{i-1}, X_i, R_i) = O_{i-1} \cap \overline{R_i} \ \cup \ X_i \cap R_i, (i > 1) \quad (2)$$

which can be further decomposed into an equivalent logic expression [10], as shown in **Fig. 6**(a):

$$\begin{cases} MUX(O_{i-1}, R_i) = O_{i-1} \cap \overline{R_i}\,, & X_i \equiv 0 \\ MUX(O_{i-1}, R_i) = O_{i-1} \cup R_i, & X_i \equiv 1 \end{cases} \quad (3)$$

This operating principle closely resembles that of a reconfigurable logic gate if the $X_i$ is the program gate signal and $O_{i-1}$ and $R_i$ are inputs [22]. The problem is that, though expression (3) is close to the NAND-NOR reconfigurable logic gate, multiple inverters are needed to make the NAND-NOR PCC work in the same way. Another 2 inverters are needed, the first one is put after the final output, and another is put before the $R_i$. Although the RFET-based NAND-NOR gate requires fewer transistors compared to a CMOS-based MUX gate, the advantage diminishes when additional inverters are included. Considering the larger footprint of a single RFET device, the overall area may show no improvement—or even result in overhead—in RFET-based PCC designs.

To address this issue, we tune the input signal by intermittently inserting inverters along the $X_i$ input port, as shown in **Fig. 6**(b), (c) and (d). Specifically, when the total chain length is odd, inverters are inserted before all odd-numbered input stages. In contrast, when the total chain length is even, inverters are inserted before all even-numbered input stages. This alternating inverter placement compensates for the polarity inversion introduced by each NAND/NOR stage, allowing the overall RFET-based chain to reproduce the same probability conversion behavior as the conventional MUX-chain PCC. This design fully exploits the compactness of RFET reconfigurable gates, replacing the conventional FinFET MUX-based logic gates that typically require 10–12 transistors. Consequently, the transistor count is significantly reduced, leading to a smaller circuit area and a shorter critical path.

**Fig. 7.** Conversion results of binary numbers with varying bit lengths under different PCC precisions. Circles denote RFET NAND-NOR PCC, squares denote MUX-chain PCC, and the dashed line indicates CMP-PCC. Curves from left to right are 3-bit to 10-bit PCCs.

A 3-bit PCC example is shown in Fig. 6(d) for better illustration. In this example, the activation input is set to binary b011, which corresponds to a decimal value of 3. The random bits are generated by the corresponding 3-bit LFSR, and for simplicity, only 8 clock cycles are considered. Based on the RFET NAND/NOR reconfigurable logic introduced earlier, with the activation bit $X_i$ acting as the PG, the three stages operate as NAND, NOR, and NOR, respectively. The output bitstream can therefore be obtained stage by stage. The final output sequence $O_3$ is b01100010, which contains three logic '1's out of eight cycles. Under unipolar stochastic encoding, this corresponds to a value of 0.375, which is equivalent to the conversion result of the conventional PCC. This example verifies that the proposed RFET-based NAND–NOR chain can achieve the same functionality as the original MUX-chain PCC.

**Lemma 1**

Consider an $N$-stage RFET NAND-NOR PCC, let

$$x_1,\ x_2, \dots, x_N \in \{0,1\},$$

be the bits of an N-bit unsigned binary input, where $x_1$ is the LSB and $x_N$ is the MSB. The corresponding integer value is

$$X = \sum_{i=1}^{N} 2^{i-1} x_i \tag{4}$$

Let $R_i$ be the random bit applied to stage $i$. The random bits are mutually independent and satisfy

$$\Pr(R_i = 0) = \Pr(R_i = 1) = \frac{1}{2} \tag{5}$$

Let $O_i$ denote the output bit of stage $i$, and let the initial input be $O_0 = 0$. Under the proposed inverter-tuning rule,

- if $N$ is even, $X_i$ is inverted for every even index $i$;
- if $N$ is odd, $X_i$ is inverted for every odd index $i$.

Then the final output probability is

$$\Pr(O_N = 1) = \begin{cases} \dfrac{X}{2^N}, N\ is\ even, \\ \dfrac{X+1}{2^N}, N\ is\ odd. \end{cases} \tag{6}$$

Therefore, for even $N$, the proposed RFET NAND–NOR PCC exactly reproduces the output probability of the conventional MUX-chain PCC. For odd $N$, it introduces only a constant positive offset of $2^{-N}$. Due to space limitations, the proof of Lemma 1 is provided in the **Supplementary Material**.

**Fig. 7** illustrates the conversion results of binary numbers to unipolar encoding SC bitstream with different bit lengths under different PCC bit widths. The circles represent the RFET NAND-NOR PCC, the squares represent the MUX-chain PCC, and the dashed line in the middle represents the CMP-PCC. It can be observed that when handling smaller odd bit lengths, the NAND-NOR PCC produces slightly higher values compared to the other two methods. This is because in equation (6), there is a term $2^{-N}$ before the summation operator.

*B. RFET-based Approximate Accumulative Parallel Counter*

APC is widely used in stochastic computing to transform stochastic signals into binary signals. Using **Fig. 8**(a) as an example [12], during each clock cycle, the APC samples the 25 input signals and counts the number of logic '1's present. The counting operation is implemented using a hierarchical adder tree composed of half adders and full adders, which generate partial sums and carry bits in parallel. These partial results are successively combined across multiple stages to accumulate the total count. The final output is a 5-bit binary number representing the total number of logic '1's among the inputs, with the maximum count of 25.

Although the APC can count the number of logic '1's in parallel without error, it consumes a large number of adders, especially when the number of inputs is high. To address this issue, the approximate accumulative parallel counter (AxPC) has been proposed [12]. In AxPC, simple logic gates are employed in the first stage to realize parity (odd–even) checking functions instead of exact addition. For example, in the referenced work [16], the first-layer full adders are replaced with NAND and NOR gates, which significantly reduces the circuit area. However, the area savings achieved by AxPC come at the cost of reduced counting accuracy.

In this work, we propose implementing and optimizing the AxPC design with RFET technology. **Fig. 8**(b) demonstrates the usage of RFET technology to build compact 25-input MAJ-AxPC. The RFET-based FA only needs two reconfigurable gates—XOR3 and MAJ3, along with a few inverters, are required. In the first stage, MAJ3 gates are used to generate the carry information, while the parity information is discarded. This structure achieves ultra-high area efficiency while maintaining considerable accuracy, as will be demonstrated in the following discussion. The architecture is particularly suitable for RFET technology because, compared with a conventional CMOS-based full adder that typically requires up to 28 transistors, the RFET-based solution achieves significant reductions in transistor count and circuit area, even when considering the larger footprint of individual RFET devices, thereby reducing energy consumption.

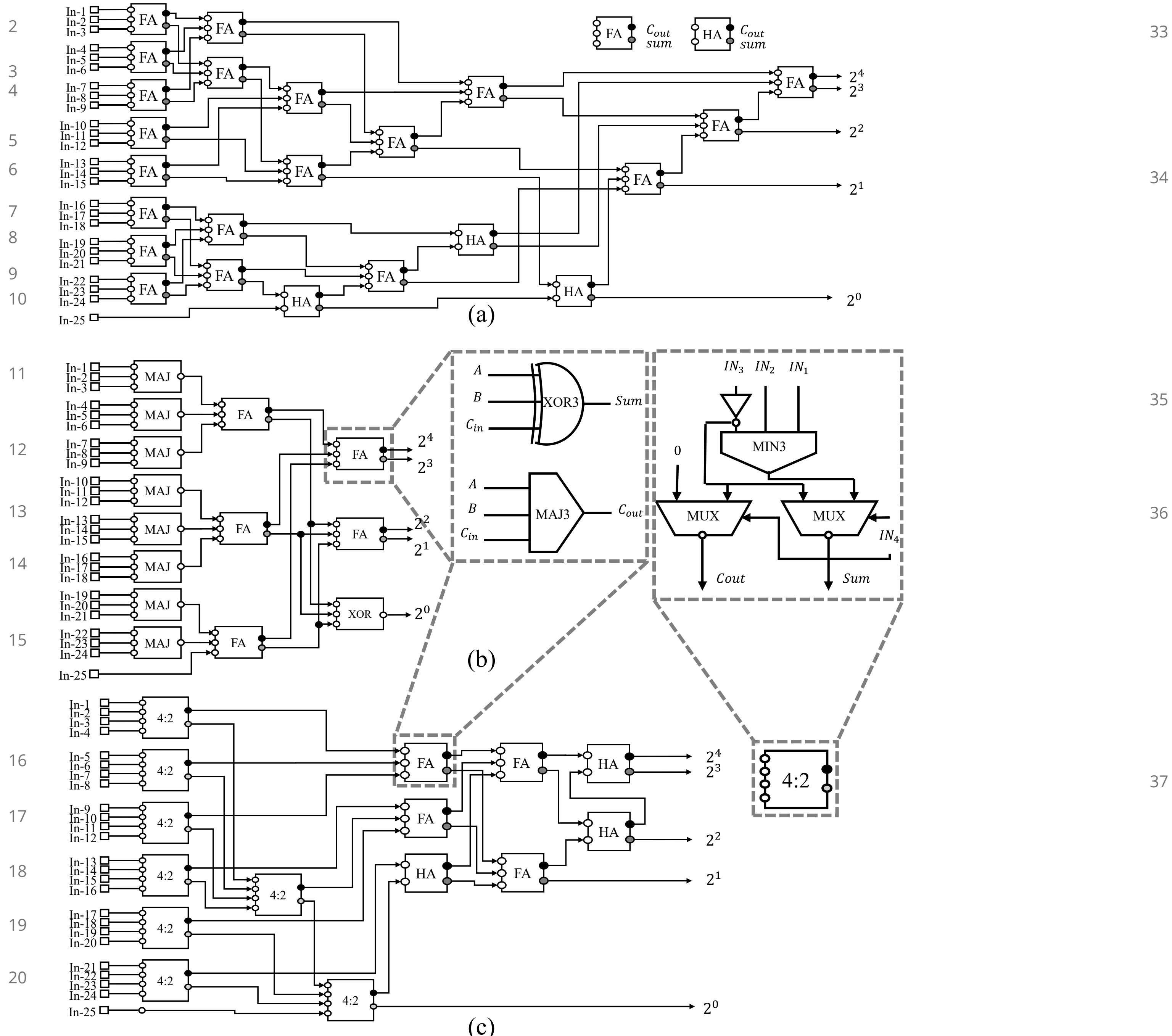

**Fig. 8.** (a) Traditional APC structure with 25 inputs. (b) Proposed RFET-based compact MAJ-AxPC structure. (c) Proposed RFET-based high accuracy COMP-AxPC structure.

Moreover, since both the referenced AxPC and the proposed MAJ-AxPC achieve area savings by sacrificing system accuracy, we propose an RFET-based 4:2 compressor AxPC (COMP-AxPC) that significantly improves accuracy while still maintaining area reduction compared with the traditional APC design. As shown in **Fig. 8**(c), the first layer of full adders is replaced by RFET-based compact 4:2 compressors [16], which will enhance the circuit's accuracy and keep the area efficient.

**Fig. 9** shows the mean squared error (MSE) and mean absolute error (MAE) as functions of the bitstream length for the referenced AxPC and proposed AxPCs, with the exact APC used as the baseline. The results are obtained from Monte Carlo simulations in which stochastic bitstreams are generated for 25 inputs, and the counter outputs are converted into stochastic outputs by a 5-bit B2S unit. For each Monte Carlo run, the output of each counter is averaged over the bitstream length. Let $y^i_{APC}$ and $y^i_{AxPC}$ denote the averaged outputs of the exact APC and an approximate APC, respectively, in the i-th run, and let N be the total number of runs. The MSE and MAE are defined as

$$MSE = \frac{1}{N}\sum_{i=1}^{N}\left(y^i_{AxPC} - y^i_{APC}\right)^2 \tag{7}$$

$$MAE = \frac{1}{N}\sum_{i=1}^{N}\left|y^i_{AxPC} - y^i_{APC}\right| \tag{8}$$

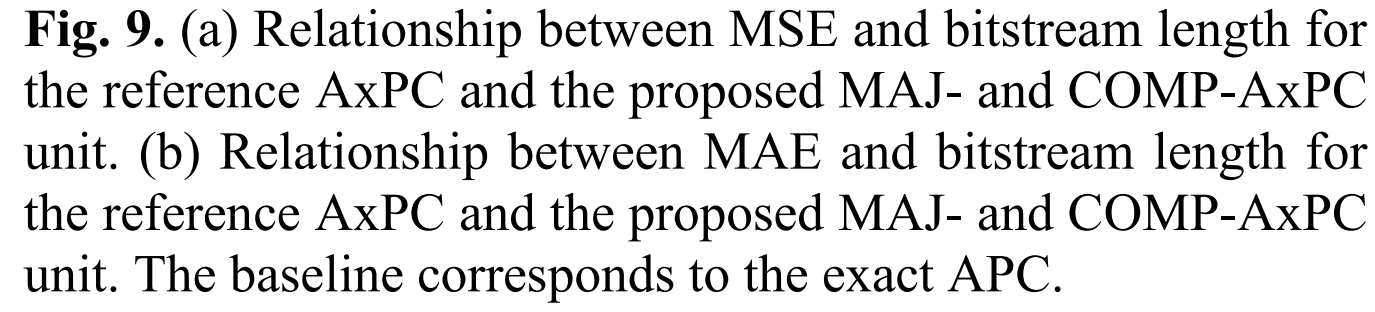

**Fig. 9.** (a) Relationship between MSE and bitstream length for the reference AxPC and the proposed MAJ- and COMP-AxPC unit. (b) Relationship between MAE and bitstream length for the reference AxPC and the proposed MAJ- and COMP-AxPC unit. The baseline corresponds to the exact APC.

The proposed MAJ-AxPC is primarily designed to achieve ultra-high area efficiency, which results in a slight sacrifice in counting accuracy. Although the proposed MAJ-AxPC exhibits slightly higher error than the traditional AxPC at shorter bitstream lengths, the error gap between the traditional AxPC and the proposed MAJ-AxPC steadily decreases as the bitstream length increases. This trend indicates that the accuracy loss introduced to improve area efficiency becomes negligible for longer bitstreams, making the proposed MAJ-AxPC sufficiently accurate and well suited for practical applications. In contrast, the proposed COMP-AxPC achieves the highest accuracy among all AxPC designs, with its MSE and MAE reduced to 12.7% and 7.5%, respectively, of those of the referenced AxPC. The hardware cost and detailed parameter comparisons are presented in the next section. In this work, we adopt the COMP-AxPC as the final solution due to its high accuracy and area efficiency when implemented using RFET technology.

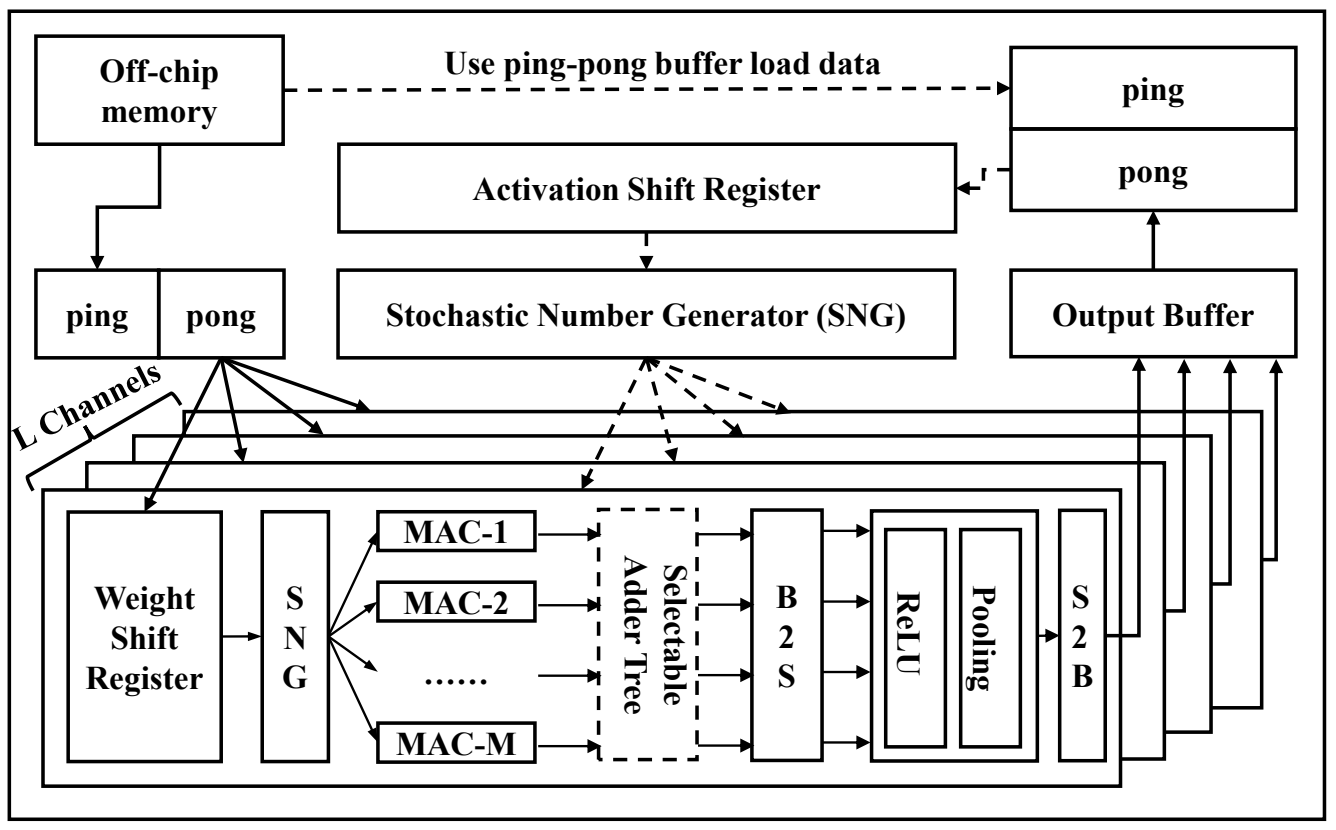

**Fig. 10.** Overall SC accelerator architecture [38].

## IV. SCNN Accelerator Architecture

The architecture presented in this section is mainly used as the evaluation framework for the proposed RFET-based PCC and AxPC, shown in **Fig. 10** [38], integrates a ping-pong buffering mechanism with activation and weight shift registers to ensure efficient data transfer from off-chip memory. In each channel, an SNG converts both activations and weights into stochastic bit streams, which are processed by 16 parallel multiply-accumulate (MAC) units to exploit parallelism. Each MAC unit comprises 25 parallel multipliers and a 25-input COMP-AxPC.

For convolutional layers, the 25-input MAC directly matches the 5×5 kernel size, allowing the configurable adder tree to be bypassed. For fully connected layers with a larger number of inputs, the same 25-input MAC/COMP-AxPC serves as a local reduction unit, while the inputs are partitioned into multiple 25-input groups. Each group is first accumulated by a local COMP-AxPC, and the resulting partial binary counts are then combined through the configurable adder tree. This hierarchical accumulation structure provides a scalable implementation while avoiding the excessive fan-in, routing congestion, and deep counter logic associated with constructing a single wide APC for hundreds of inputs.

If sufficient computation channels and memory bandwidth are available, multiple MAC/AxPC groups can operate in parallel to improve throughput. When the fully connected layer exceeds the available on-chip resources, the same MAC/AxPC units are reused across multiple input groups, and the partial sums are accumulated over multiple cycles. This time-multiplexed strategy trades increased latency for reduced hardware cost and routing complexity. Subsequently, a B2S converter is employed to enable transformations and preserve compatibility within the stochastic domain. After computation, the results pass through an optional ReLU/MP stage and a stochastic-to-binary (S2B) converter that converts outputs back to binary representation [39]. The final results are stored in the output buffer.

This architecture can be modified for different channel numbers, making it adaptable to various application scenarios.

TABLE I

ACCURACY, AREA, DELAY, AND ENERGY COMPARISON BETWEEN FINFET- AND RFET-BASED 8-BIT PCC AND 25-INPUT 32-BIT APCs

| | 8-bit PCC* | | | Traditional APC | | | Reference AxPC | | | Proposed MAJ-AxPC | | | Proposed COMP-AxPC | | |
|---|---|---|---|---|---|---|---|---|---|---|---|---|---|---|---|
| **MSE/MAE** | NA | | | As the baseline | | | 0.0059/0.059 (medium) | | | 0.0067/0.064 (high) | | | 0.0051/0.054 (low) | | |
| **Technology** | **Area ($\mu m^2$)** | **Delay (ps)** | **Energy (fJ)** | **Area ($\mu m^2$)** | **Delay (ps)** | **Energy (fJ)** | **Area ($\mu m^2$)** | **Delay (ps)** | **Energy (fJ)** | **Area ($\mu m^2$)** | **Delay (ps)** | **Energy (fJ)** | **Area ($\mu m^2$)** | **Delay (ps)** | **Energy (fJ)** |
| **FinFET10nm** | 2.2 | 242 | 4.1 | 32.1 | 352 | 87.6 | 10.7 | 189 | 21.8 | 7.3 | 166 | 19.3 | 13.8 | 269 | 33.5 |
| **RFET10nm** | 2.0 | 142 | 2.9 | 26.2 | 334 | 54.6 | 9.1 | 124 | 16.1 | 6.0 | 151 | 12.0 | 12.1 | 171 | 20.9 |
| **RFET Gain** | 9.1% | 41.3% | 29.3% | 18.4% | 5.7% | 37.7% | 15.0% | 34.4% | 26.1% | 17.8% | 9.0% | 37.8% | 12.3% | 36.4% | 37.6% |

* The RFET-based 8-bit PCC is implemented using the proposed NAND–NOR architecture, whereas the FinFET-based 8-bit PCC employs the conventional MUX-chain architecture.

The system is based on 8-bit accuracy, with weights and activations loaded from off-chip GDDR5 memory operating at 7000 MHz, providing a loading speed of approximately 224 B/ns. Due to page limitations, details of the pipeline strategy are provided in the **Supplementary Material**.

## V. SIMULATION RESULTS ANALYSIS

This work focuses on comparing SCNN accelerators implemented under an EDA-based synthesis flow using RFET standard-cell libraries. While this methodology does not correspond to a fabricated silicon prototype, it provides a practical and widely adopted approach for evaluating emerging device technologies at the architectural level, especially given the current limitations in RFET manufacturing maturity and the high cost associated with advanced-node tape-out.

Based on this evaluation framework, this section presents the simulation results of the proposed SCNN accelerators across different technologies. We focus on exploring the potential of RFETs in stochastic computing, with FinFETs serving as the reference counterpart for comparison. System-level simulations are performed using the Cadence Genus EDA tool with suitable constraints, based on the open-source 10nm three-independent-gate 4-nanowire RFET standard cell library developed by Gauchi et al [40]. Since no corresponding open-source 10nm FinFET library is available, we adopt the widely used ASAP 7nm library and scale it to the equivalent 10nm node [41]. The scaling method used here is supported by [42, 43].

To ensure a fair comparison, both RFET and FinFET use the same logic gate library with minimum drive strength. This enforces a consistent cell-selection policy across both technologies and avoids introducing bias from different drive-strength optimization strategies. Although industrial designs typically perform drive-strength sizing along critical paths, this is not feasible here due to the lack of comparable multi-drive RFET and FinFET libraries. Regarding supply voltage, RFET and FinFET are characterized at 0.85 V and 0.7 V, respectively, following their available library conditions and typical energy-delay product (EDP) operating points. The higher RFET voltage is mainly due to its lower drive current, which requires higher voltage for competitive speed

TABLE II

CHANNEL-LEVEL AREA, DELAY, AND ENERGY COMPARISON

| Parameters | FinFET | RFET | Gain |
|---|---|---|---|
| **Channel Area ($\mu m^2$)** | 2710 | 2223 | 18% |
| **Min Clock Period (ns)** | 0.82 | 0.64 | 22% |
| **Switching Energy (pJ)** | 4.22 | 2.94 | 30% |

For workload evaluation, we focus on two commonly used CNN datasets, MNIST and CIFAR-10, which have been widely adopted in prior SCNN accelerator studies. This choice provides mature benchmarks and well-established reference results for fair comparison of different hardware designs, while avoiding additional uncertainty from less standardized workloads. This allows us to clearly evaluate the benefits of the proposed RFET-based components.

### *A. Performance Comparison of FinFET and RFET Blocks*

As shown in Table I, two key components—an 8-bit PCC and the 25-input COMP-AxPC with 32-bit bitstream—are analyzed in detail under both FinFET and RFET technologies. For the PCC, the RFET-based implementation demonstrates notable improvements in area, delay, and switching energy compared with its FinFET counterpart, achieving reductions of approximately 9%, 41%, and 29%, respectively. The performance improvements mainly originate from the simplified NAND–NOR logic structure adopted by the RFET-based PCC, which replaces the conventional MUX-chain architecture used in the FinFET implementation. This architectural simplification substantially reduces the required transistor count, leading to a smaller circuit area despite the larger footprint of individual RFET devices. Moreover, the lower internal and load capacitances along the critical path compensate for the lower RFET on-state current, resulting in a shorter propagation delay. The reduced transistor count also decreases the switching capacitance, thereby lowering the switching energy.

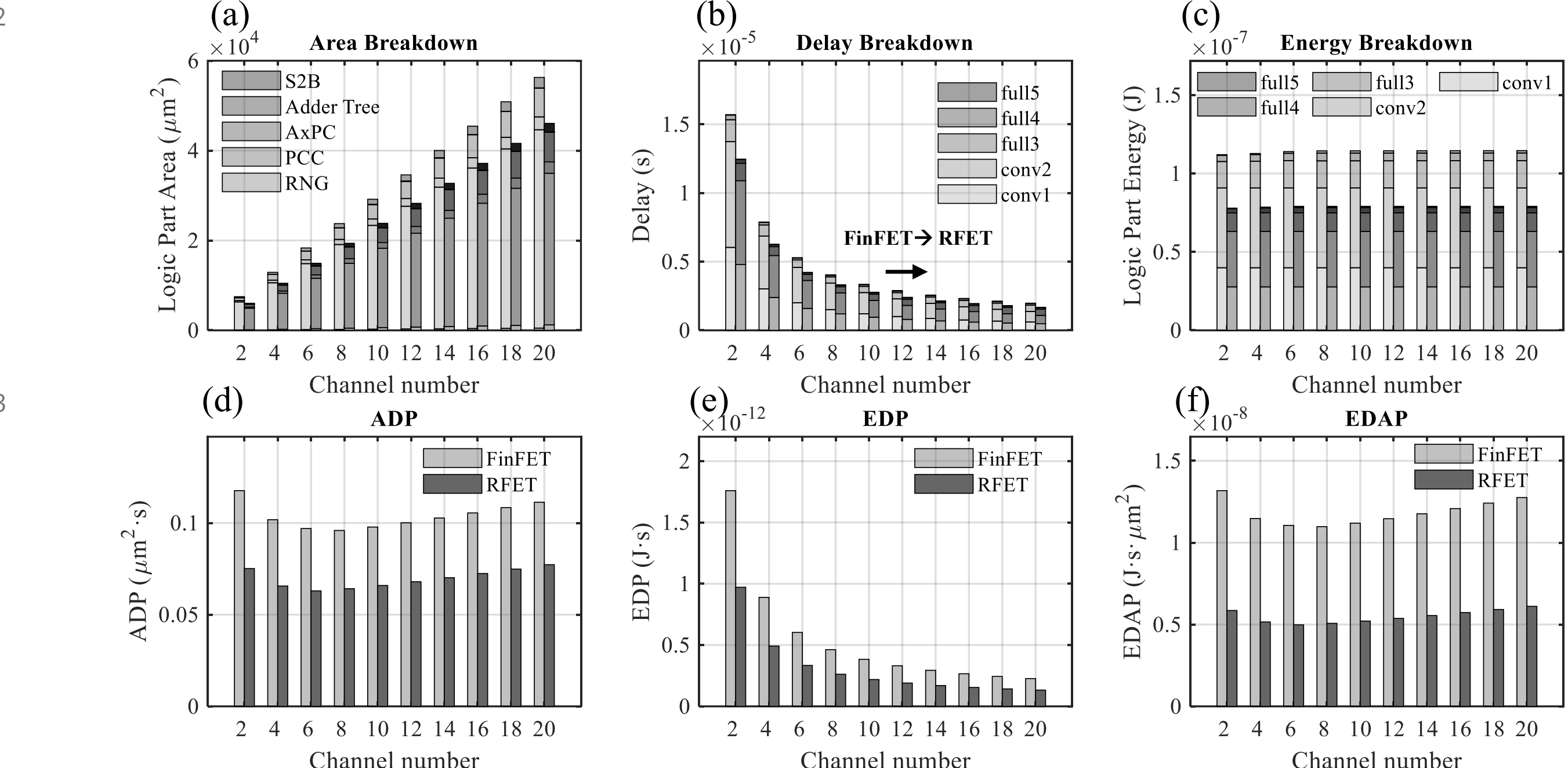

**Fig. 11.** System-level performance comparison between FinFET- and RFET-based SCNNs. Left bars represent FinFET, right bars represent RFET.

A similar performance trend is observed for the RFET-based APC and AxPCs. In most scenarios, RFETs can effectively optimize performance compared to FinFETs. The proposed RFET-based MAJ-AxPCs achieve high performance while incurring only a slight loss in accuracy, whereas the COMP-AxPC attains higher accuracy while preserving the performance advantages over traditional FinFET-based APC and AxPC designs. Table II compares FinFET and RFET implementations in terms of area, clock period, and energy per channel. The RFET implementation achieves an 18% reduction in channel area, a 22% improvement in clock period, and a 30% decrease in switching energy, clearly demonstrating its advantages over FinFET technology.

### *B. System-level Simulation Results*

#### *1) Overall Performance Compared with FinFET*

The system-level simulation is conducted to identify the optimal channel number and to compare the performance of FinFET-based and RFET-based accelerators. As shown in **Fig. 11**, the logic part area of the system increases linearly with the number of channels, while the latency decreases as the channel count grows. The area breakdown indicates that the primary contributor to the overall area is the PCC, with the AxPC and adder tree also contributing significantly. With more channels, additional resources are available for parallel processing; however, the delay eventually approaches a limit imposed by memory bandwidth. Most of the latency originates from the convolutional layers, which involve a large number of neurons. The energy consumption of the logic part remains relatively unchanged, as the majority of the energy arises from switching activity, and the total switching-induced energy remains constant.

To determine an appropriate channel number, three commonly used system-level metrics—area-delay product (ADP), EDP, and energy-delay-area product (EDAP)—are employed as evaluation criteria. Based on the plots of ADP and EDAP, the optimal channel number is found to be 8 for both FinFET and RFET technologies. At this configuration, the RFET-based accelerator achieves reductions of up to 18% in area, 18% in delay, and 31% in energy, resulting in an overall EDAP improvement of 54%.

#### *2) Impact of Bitstream and System Precision on Accuracy*

To determine the optimal bitstream length for the SCNN system, training is performed using PyTorch. However, since PyTorch does not natively support all the mathematical functions required for SCNNs, the training process relies on the insertion of equivalent SC models. Specifically, the mathematical model of SC is encapsulated as a Python function and integrated into the training pipeline. The inference flow is constructed in the same manner, strictly adhering to the adopted SCNN structure.

Two commonly used CNN datasets, MNIST and CIFAR-10, are employed for evaluation. The MNIST is based on the LeNet-5 CNN structure, while the CIFAR-10 follows the same network structure as the reference works [36, 38]. Overall, as shown in the **Fig. 12**, model accuracy increases with both the bitstream length and the system precision. The system precision defines the accuracy ceiling; higher precision leads to a higher upper limit, although the improvement becomes negligible once the precision exceeds five bits. With respect to bitstream length, accuracy improves rapidly when the length is small, but the rate of improvement slows down as the length increases, eventually reaching a stable level. Based on the simulation results, the

TABLE III

COMPARISON WITH OTHER STATE-OF-THE-ART ACCELERATOR WORKS

| | **ISSCC 21 [44]** | **TCAD 18 [6]** | **TCASII 22 [45]** | **SSCL 22 [38]** | **TNNLS 23 [36]** | **This Work** | |
|---|---|---|---|---|---|---|---|
| **NN Type** | Digital | SC | SC | SC | SC | SC | |
| **Impl. Type** | Silicon | Silicon | Synthesis | Silicon | Synthesis/FPGA | Synthesis | |
| **Technology** | CMOS | CMOS | CMOS | CMOS | CMOS | CMOS | RFET |
| **Tech Node** | 7nm | 45nm | 65nm | 14nm | 40nm | 10nm | 10nm |
| **Voltage** | 0.55-0.75V | Not Applied | 1.0V | 0.6-0.9V | Not Applied | 0.7V | 0.85V |
| **Clock** | 1.0-1.6GHz | 481MHz | 909MHz | 250-500MHz | 200 MHz | 1.25GHz | 1.57GHz |
| **Precision** | FP8-32 | 7 | 5 | 4-8 | 8 | 8 | 8 |
| **Area** | 19.6mm² | 22.9mm² | 0.006mm² | 0.5mm² | 2.1mm² | 0.237mm² | 0.194mm² |
| **Power** | Not Applied | 2600mW | 4.06mW | 16-68mW | 651mW | 13.9mW | 12.2mW |
| **TOPS/W** | 8.9-16.5 | 5.66 | 2.17 | 4.4-75 | 0.34 | 12.12 | 16.82 |
| **TOPS/mm2** | 3.27-5.22 | 0.64 | 1.44 | 0.3-4.8 | 0.11 | 7.09 | 10.60 |
| **MNIST Acc.** | Not Applied | 99.07% | 98.8% | 95.1%-98.7% | 97.6% | 98.5% | |
| **CIFAR-10 Acc.** | Not Applied | Not Applied | Not Applied | 69.4%-71.7% | 81% | 73.1% | |

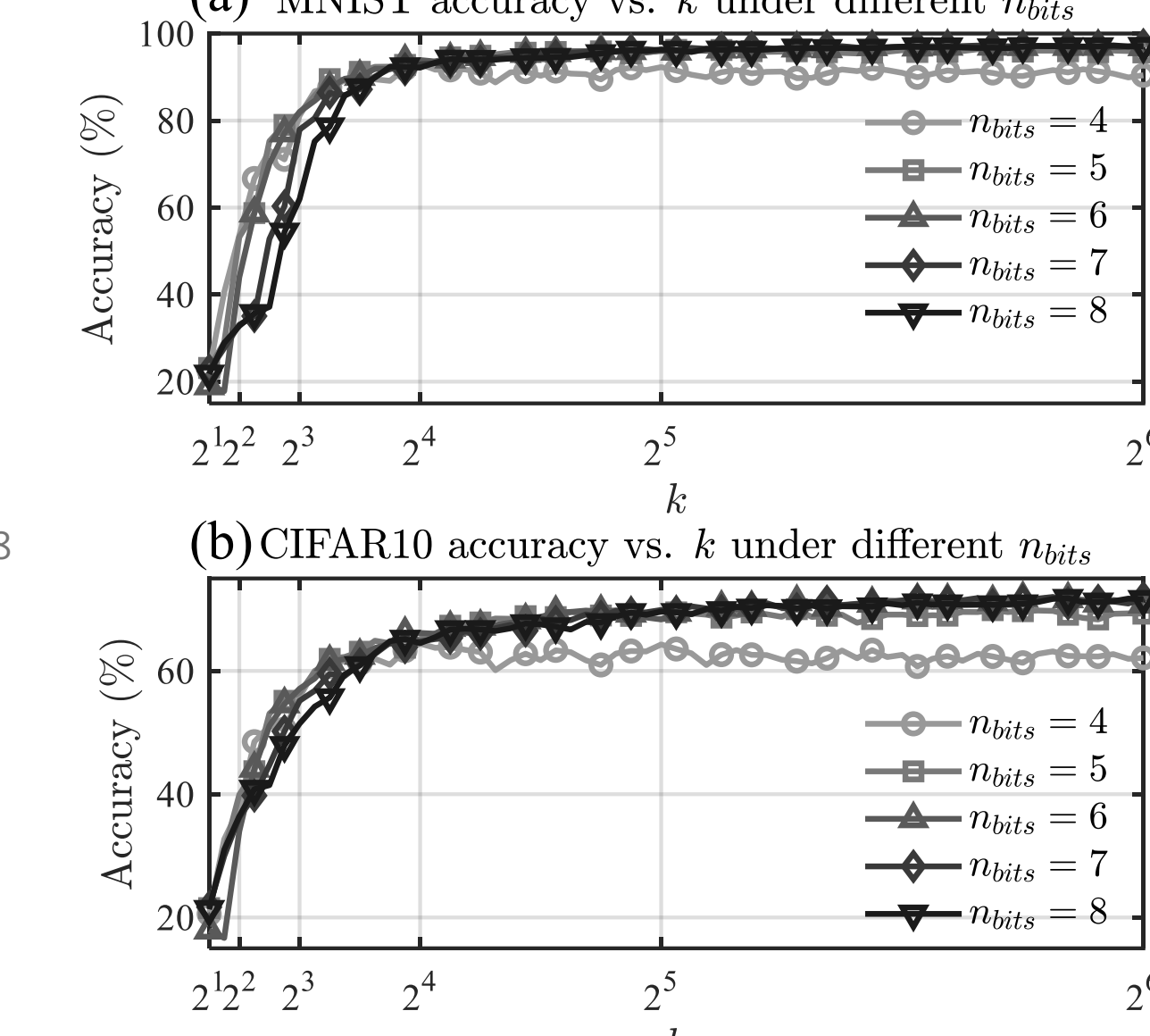

**Fig. 12.** Relationship between accuracy and bitstream length under varying system precision levels.

SCNN model achieves accuracies of up to 98.5% on MNIST and 73.1% on CIFAR-10.

### C. Comparison with State-of-the-Art Works

Table III provides a comparative overview of several recent works on SCNN and digital neural network accelerators [6, 36, 38, 44, 45]. Despite the different implementation platforms, the comparison is intended only as a high-level reference to representative state-of-the-art works. The design presented in this work employs RFET-based technology at the 10nm node, and the SCNN operates with 8-bit system precision and uses 32-bit bitstream length with unipolar encoding. Operating at 0.85V with a maximum clock frequency of 1.57GHz, the design maintains competitive performance among SC implementations. It also features a compact area footprint of 0.194mm², including 10KB on-chip memory, which is smaller than that of designs such as SSCL 22 (0.5mm²) and TNNLS 23 (2.1mm²). The simulated power consumption is at 12.2mW, contributing to an energy efficiency of 16.82 TOPS/W, which compares favorably with several prior works. Additionally, a computational density of 10.6 TOPS/mm² is achieved. In terms of inference accuracy, the design attains 98.5% on the MNIST and 73.1% on the CIFAR-10 datasets, demonstrating reasonable performance for SC-based approaches. To improve TOPS/W, shorter bitstreams such as 16-bit or even 8-bit representations can be used. However, this comes at the cost of reduced numerical precision and degraded model accuracy.

### D. Future Work

Future work will extend the proposed RFET-based stochastic computing framework in three directions. First, we will investigate its application to larger AI models, such as transformer-based architectures and large language models (LLMs), with a focus on efficient stochastic implementations of transformer-specific operations. Second, we will explore RFET-based bipolar stochastic computing by leveraging the efficient XOR/XNOR logic enabled by RFET reconfigurability to support signed computations more efficiently. Finally, we will develop an FPGA-based prototype for system-level functional validation and large-scale inference evaluation, providing a practical step toward future RFET hardware implementation.

## VI. CONCLUSIONS

In this paper, a compact and efficient RFET-based SCNN accelerator is proposed, featuring optimized key components, such as PCCs and APCs. Leveraging the reconfigurable properties of RFETs, we demonstrate—through both mathematical analysis and simulation—that PCCs can be designed with significantly reduced area. System-level simulations based on a widely adopted accelerator architecture indicate that, under the same technology node, the RFET-based SCNN achieves up to 18% area reduction, 18% latency improvement, and 31% switching energy savings under 8-bit accuracy and 32-bit bitstream compared to the FinFET counterpart. These enhancements lead to an overall improvement of 39% in TOPS/W and 49% in TOPS/mm².

## References

[1] N. K. Upadhyay *et al.*, "Emerging memory devices for neuromorphic computing," *Advanced Materials Technologies,* vol. 4, no. 4, p. 1800589, 2019.

[2] O. I. Abiodun *et al.*, "State-of-the-art in artificial neural network applications: A survey," *Heliyon,* vol. 4, no. 11, 2018.

[3] Y. Liu *et al.*, "A survey of stochastic computing neural networks for machine learning applications," *IEEE Transactions on Neural Networks and Learning Systems,* vol. 32, no. 7, pp. 2809-2824, 2020.

[4] A. Ardakani *et al.*, "VLSI implementation of deep neural network using integral stochastic computing," *IEEE Transactions on Very Large Scale Integration (VLSI) Systems,* vol. 25, no. 10, pp. 2688-2699, 2017.

[5] A. Ren *et al.*, "Sc-dcnn: Highly-scalable deep convolutional neural network using stochastic computing," *ACM Sigplan Notices,* vol. 52, no. 4, pp. 405-418, 2017.

[6] Z. Li *et al.*, "HEIF: Highly efficient stochastic computing-based inference framework for deep neural networks," *IEEE Transactions on Computer-Aided Design of Integrated Circuits and Systems,* vol. 38, no. 8, pp. 1543-1556, 2018.

[7] K. Kim *et al.*, "An energy-efficient random number generator for stochastic circuits," in *2016 21st Asia and South Pacific Design Automation Conference (ASP-DAC)*, 2016: IEEE, pp. 256-261.

[8] K. Zhong *et al.*, "A Brief Survey on Randomizer Design and Optimization for Efficient Stochastic Computing," in *2024 IEEE International Test Conference in Asia (ITC-Asia)*, 2024: IEEE, pp. 1-6.

[9] K. Zhong *et al.*, "Towards low-cost high-accuracy stochastic computing architecture for univariate functions: design and design space exploration," in *2022 Design, Automation & Test in Europe Conference & Exhibition (DATE)*, 2022: IEEE, pp. 346-351.

[10] Y. Ding *et al.*, "Generating multiple correlated probabilities for MUX-based stochastic computing architecture," in *2014 IEEE/ACM International Conference on Computer-Aided Design (ICCAD)*, 2014: IEEE, pp. 519-526.

[11] C. Collinsworth *et al.*, "Stochastic number generators with minimum probability conversion circuits," in *2021 IEEE Computer Society Annual Symposium on VLSI (ISVLSI)*, 2021: IEEE, pp. 49-54.

[12] K. Kim *et al.*, "Approximate de-randomizer for stochastic circuits," in *2015 International SoC Design Conference (ISOCC)*, 2015: IEEE, pp. 123-124.

[13] C. Pan *et al.*, "A proposal for energy-efficient cellular neural network based on spintronic devices," *IEEE Transactions on Nanotechnology,* vol. 15, no. 5, pp. 820-827, 2016.

[14] M. W. Daniels *et al.*, "Energy-efficient stochastic computing with superparamagnetic tunnel junctions," *Physical review applied,* vol. 13, no. 3, p. 034016, 2020.

[15] C. Pan *et al.*, "An expanded benchmarking of beyond-CMOS devices based on Boolean and neuromorphic representative circuits," *IEEE Journal on Exploratory Solid-State Computational Devices and Circuits,* vol. 3, pp. 101-110, 2018.

[16] N. Kavand *et al.*, "Design of energy-efficient RFET-based exact and approximate 4: 2 compressors and multipliers," *IEEE Transactions on Circuits and Systems II: Express Briefs,* vol. 70, no. 9, pp. 3644-3648, 2023.

[17] Q. Qu *et al.*, "Fast and Energy-Efficient Analog Accelerator for Vision Transformer," in *2025 IEEE 68th International Midwest Symposium on Circuits and Systems (MWSCAS)*, 2025: IEEE, pp. 827-831.

[18] A. Heinzig *et al.*, "Reconfigurable silicon nanowire transistors," *Nano letters,* vol. 12, no. 1, pp. 119-124, 2012.

[19] M. Reuter *et al.*, "From mosfets to ambipolar transistors: Standard cell synthesis for the planar rfet technology," *IEEE Transactions on Circuits and Systems I: Regular Papers,* vol. 68, no. 1, pp. 114-125, 2020.

[20] T. Mikolajick *et al.*, "Reconfigurable field effect transistors: A technology enablers perspective," *Solid-State Electronics,* vol. 194, p. 108381, 2022.

[21] T. Mikolajick *et al.*, "The RFET—A reconfigurable nanowire transistor and its application to novel electronic circuits and systems," *Semiconductor Science and Technology,* vol. 32, no. 4, p. 043001, 2017.

[22] S. Rai *et al.*, "Designing efficient circuits based on runtime-reconfigurable field-effect transistors," *IEEE Transactions on Very Large Scale Integration (VLSI) Systems,* vol. 27, no. 3, pp. 560-572, 2018.

[23] J. Romero-González *et al.*, "An efficient adder architecture with three-independent-gate field-effect transistors," in *2018 IEEE International Conference on Rebooting Computing (ICRC)*, 2018: IEEE, pp. 1-8.

[24] J. Trommer *et al.*, "Functionality-enhanced logic gate design enabled by symmetrical reconfigurable silicon nanowire transistors," *IEEE Transactions on Nanotechnology,* vol. 14, no. 4, pp. 689-698, 2015.

[25] S. Lu *et al.*, "A Novel RFET-Based FPGA Architecture Based on Delay-Aware Packing Algorithm," *IEEE Transactions on Emerging Topics in Computing,* 2025.

[26] S. Lu *et al.*, "Technology/System Co-Optimization for FPGA Using Emerging Reconfigurable Logic Device," *ACM Journal on Emerging Technologies in Computing Systems,* vol. 21, no. 3, pp. 1-21, 2025.

[27] R. Saravanan *et al.*, "Reconfigurable fet approximate computing-based accelerator for deep learning applications," in *2023 IEEE International Symposium on Circuits and Systems (ISCAS)*, 2023: IEEE, pp. 1-5.

[28] J. Zhang *et al.*, "Configurable circuits featuring dual-threshold-voltage design with three-independent-gate silicon nanowire FETs," *IEEE Transactions on Circuits and Systems I: Regular Papers,* vol. 61, no. 10, pp. 2851-2861, 2014.

[29] M. Simon *et al.*, "A wired-AND transistor: Polarity controllable FET with multiple inputs," in *2018 76th Device Research Conference (DRC)*, 2018: IEEE, pp. 1-2.

[30] J. Trommer *et al.*, "Enabling energy efficiency and polarity control in germanium nanowire transistors by individually gated nanojunctions," *ACS nano,* vol. 11, no. 2, pp. 1704-1711, 2017.

[31] J. Zhang *et al.*, "Polarity-controllable silicon nanowire transistors with dual threshold voltages," *IEEE Transactions on Electron Devices,* vol. 61, no. 11, pp. 3654-3660, 2014.

[32] J.-H. Bae *et al.*, "Reconfigurable field-effect transistor as a synaptic device for XNOR binary neural network," *IEEE Electron Device Letters,* vol. 40, no. 4, pp. 624-627, 2019.

[33] A. Alaghi *et al.*, "The promise and challenge of stochastic computing," *IEEE Transactions on Computer-Aided Design of Integrated Circuits and Systems,* vol. 37, no. 8, pp. 1515-1531, 2017.

[34] P. Purwono *et al.*, "Understanding of convolutional neural network (cnn): A review," *International Journal of Robotics and Control Systems,* vol. 2, no. 4, pp. 739-748, 2022.

[35] D. Ghimire *et al.*, "A survey on efficient convolutional neural networks and hardware acceleration," *Electronics,* vol. 11, no. 6, p. 945, 2022.

[36] C. F. Frasser *et al.*, "Fully parallel stochastic computing hardware implementation of convolutional neural networks for edge computing applications," *IEEE Transactions on Neural Networks and Learning Systems,* vol. 34, no. 12, pp. 10408-10418, 2022.

[37] J. Yang *et al.*, "A 65-nm Digital Stochastic Compute-in-Memory CNN Processor With 8-bit Precision," *IEEE Journal of Solid-State Circuits,* 2025.

[38] W. Romaszkan *et al.*, "A 4.4–75-tops/w 14-nm programmable, performance-and precision-tunable all-digital stochastic computing neural network inference accelerator," *IEEE Solid-State Circuits Letters,* vol. 5, pp. 206-209, 2022.

[39] H. Xiong *et al.*, "Hardware implementation of an improved stochastic computing based deep neural network using short sequence length," *IEEE Transactions on Circuits and Systems II: Express Briefs,* vol. 67, no. 11, pp. 2667-2671, 2020.

[40] R. Gauchi *et al.*, "An open-source three-independent-gate fet standard cell library for mixed logic synthesis," in *2022 IEEE International Symposium on Circuits and Systems (ISCAS)*, 2022: IEEE, pp. 273-277.

[41] L. T. Clark *et al.*, "ASAP7: A 7-nm finFET predictive process design kit," *Microelectronics Journal,* vol. 53, pp. 105-115, 2016.

[42] I. I. R. f. Devices *et al.* "International Roadmap for Devices and Systems (IRDS™) 2017 Edition." IEEE. https://irds.ieee.org/editions/2017 (accessed October 5, 2025).

[43] M. N. A. Taj, *Intel Chip Manufacturing Technology Roadmap*. 2022.

[44] A. Agrawal *et al.*, "9.1 A 7nm 4-core AI chip with 25.6 TFLOPS hybrid FP8 training, 102.4 TOPS INT4 inference and workload-aware throttling," in *2021 IEEE International Solid-State Circuits Conference (ISSCC)*, 2021, vol. 64: IEEE, pp. 144-146.

[45] H. Wang *et al.*, "An efficient stochastic convolution architecture based on fast FIR algorithm," *IEEE Transactions on Circuits and Systems II: Express Briefs,* vol. 69, no. 3, pp. 984-988, 2021.

# Supplementary Material

## I. PROOF OF LEMMA 1

Let $N \geq 1$ be the PCC precision. Let $x_1, x_2, \dots, x_N \in \{0,1\}$ be the bits of an $N$-bit unsigned binary input $X$, where $x_1$ is the least significant bit (LSB) and $x_N$ is the most significant bit (MSB). Therefore,

$$X = \sum_{i=1}^{N} 2^{i-1} x_i. \tag{1}$$

Let $R_i \in \{0,1\}$ be the random bit applied to stage $i$. Each random bit has equal probabilities of being 0 or 1:

$$\Pr(R_i = 0) = \Pr(R_i = 1) = \frac{1}{2}. \tag{2}$$

The random bits $R_1, R_2, \dots, R_N$ are mutually independent.
Let $O_i \in \{0,1\}$ be the output of stage $i$, and let $O_0 = 0$ be the fixed input of the first stage. Let $T_i \in \{0,1\}$ be the programming bit actually applied to stage $i$ after inverter tuning. Here, $T_i = 0$ selects the NAND function, whereas $T_i = 1$ selects the NOR function.
According to the inverter-tuning rule, when $N$ is even,

$$T_i = \begin{cases} x_i, & i \text{ odd}, \\ 1 - x_i, & i \text{ even}, \end{cases} \qquad N \text{ even}, \tag{3}$$

and when $N$ is odd,

$$T_i = \begin{cases} 1 - x_i, & i \text{ odd}, \\ x_i, & i \text{ even}. \end{cases} \qquad N \text{ odd}. \tag{4}$$

For two binary inputs $a, b \in \{0,1\}$,

$$\begin{aligned} \text{NAND}(a,b) &= 1 - ab, \\ \text{NOR}(a,b) &= (1-a)(1-b) \end{aligned} \tag{5}$$

Hence, the output of stage $i$ can be written as

$$\begin{aligned} O_i = (1 - T_i)&\text{NAND}(O_{i-1}, R_i) \\ +T_i&\text{NOR}(O_{i-1}, R_i), \qquad 1 \leq i \leq N. \end{aligned} \tag{6}$$

Since $T_i$ is binary, only one of the two logic functions is selected at each stage.
Now define

$$q_i = \Pr(O_i = 1), \qquad q_0 = 0. \tag{7}$$

Because $O_{i-1}$ depends only on the previous random bits $R_1, \dots, R_{i-1}$, while $R_i$ is independent of them, $O_{i-1}$ and $R_i$ are independent.
For the NAND case, the output is 0 only when both inputs are 1. Therefore,

$$\begin{aligned} &\Pr(\text{NAND}(O_{i-1}, R_i) = 1) \\ &= 1 - \Pr(O_{i-1} = 1, R_i = 1) \\ &= 1 - \Pr(O_{i-1} = 1)\Pr(R_i = 1) \\ &= 1 - \frac{q_{i-1}}{2}. \end{aligned} \tag{8}$$

For the NOR case, the output is 1 only when both inputs are 0. Therefore,

$$\begin{aligned} &\Pr(\text{NOR}(O_{i-1}, R_i) = 1) \\ &= \Pr(O_{i-1} = 0, R_i = 0) \\ &= \Pr(O_{i-1} = 0)\Pr(R_i = 0) \\ &= \frac{1 - q_{i-1}}{2}. \end{aligned} \tag{9}$$

Using (6), the output probability of stage $i$ is

$$\begin{aligned} q_i &= (1 - T_i)\left(1 - \frac{q_{i-1}}{2}\right) + T_i\left(\frac{1 - q_{i-1}}{2}\right) \\ &= -\frac{q_{i-1}}{2} + 1 - \frac{T_i}{2}. \end{aligned} \tag{10}$$

Starting from $q_0 = 0$, repeated substitution of (10) gives

$$\begin{aligned} q_N = &\left(-\frac{1}{2}\right)^{N-1}\left(1 - \frac{T_1}{2}\right) + \left(-\frac{1}{2}\right)^{N-2}\left(1 - \frac{T_2}{2}\right) \\ &+ \cdots + \left(-\frac{1}{2}\right)\left(1 - \frac{T_{N-1}}{2}\right) + \left(1 - \frac{T_N}{2}\right). \end{aligned} \tag{11}$$

Equivalently,

$$q_N = \sum_{i=1}^{N} \left(-\frac{1}{2}\right)^{N-i}\left(1 - \frac{T_i}{2}\right). \tag{12}$$

We next examine the contribution of each input bit $x_i$. There are two cases. If $N - i$ is even, then $i$ and $N$ have the same parity. From (3) and (4),

$$T_i = 1 - x_i. \tag{13}$$

The corresponding term in (12) becomes

$$\left(-\frac{1}{2}\right)^{N-i}\left(1-\frac{T_i}{2}\right)=\frac{1}{2^{N-i}}\left(1-\frac{1-x_i}{2}\right) = \frac{1+x_i}{2^{N-i+1}}. \tag{14}$$

If $N-i$ is odd, then $i$ and $N$ have opposite parity. In this case,

$$T_i = x_i. \tag{15}$$

The corresponding term becomes

$$\left(-\frac{1}{2}\right)^{N-i}\left(1-\frac{T_i}{2}\right)=-\frac{1}{2^{N-i}}\left(1-\frac{x_i}{2}\right) = -\frac{1}{2^{N-i}}+\frac{x_i}{2^{N-i+1}}. \tag{16}$$

From (13) and (14), the coefficient of $x_i$ is the same in both cases:

$$\frac{1}{2^{N-i+1}}=\frac{2^{i-1}}{2^N}. \tag{17}$$

Therefore, the total contribution of all input bits is

$$\sum_{i=1}^{N}\frac{2^{i-1}x_i}{2^N}=\frac{1}{2^N}\sum_{i=1}^{N}2^{i-1}\,x_i = \frac{X}{2^N}. \tag{18}$$

It remains to evaluate the constant terms. From (13) and (14), whenever two successive exponents are $2j$ and $2j+1$, their constant parts cancel:

$$\frac{1}{2^{2j+1}}-\frac{1}{2^{2j+1}}=0. \tag{19}$$

When $N$ is even, all constant terms can be grouped into such pairs. Therefore, no constant term remains, and

$$\Pr(O_N=1)=\frac{X}{2^N},\qquad N \text{ even}. \tag{20}$$

When $N$ is odd, all paired constant terms still cancel, but one additional term remains. This unpaired term corresponds to $N-i=N-1$ and has value

$$\frac{1}{2^N} \tag{21}$$

Therefore,

$$\Pr(O_N=1)=\frac{X}{2^N}+\frac{1}{2^N}=\frac{X+1}{2^N},\qquad N \text{ odd}. \tag{22}$$

Combining the even- and odd-$N$ cases,

$$\Pr(O_N=1)=\begin{cases}\dfrac{X}{2^N}, & N \text{ even},\\ \dfrac{X+1}{2^N}, & N \text{ odd}.\end{cases} \tag{23}$$

This completes the proof.

## II. PIPELINE STRATEGY FOR OFF-CHIP MEMORY BANDWIDTH LIMIT

Based on the proposed architecture, the off-chip memory bandwidth can become a major bottleneck because the number of neurons physically instantiated on chip may exceed the number of neurons whose activation and weight data can be supplied by the memory interface within the required time interval. To characterize this mismatch, let $n_{\text{onchip}}$ denote the number of neurons available on chip for the current layer and $n_{\text{memcover}}$ denote the number of neurons that can be supported by one memory-loading opportunity under the available off-chip bandwidth. In addition, $k$ denotes the stochastic bitstream length and $\tau$ denotes the system clock period. The key idea of the proposed scheduling strategy is to pipeline the processing of different stochastic bit positions, such that data loading for one group of neurons can overlap with the computation of other groups. In this way, the same on-chip computing resources can be reused across different bit positions while reducing idle intervals caused by insufficient off-chip data supply.

As summarized in Algorithm 1 and illustrated in Fig. 1, three operating cases are considered. When $n_{\text{onchip}} < n_{\text{memcover}}$, the memory interface is able to provide sufficient data to activate all on-chip neurons required by the current layer. Therefore, no inter-bit pipelining is necessary, as shown in Fig. 1(a). All available neurons can operate in parallel, i.e., $n_{\text{parallel}} = n_{\text{onchip}}$, and the $k$ stochastic bits are processed sequentially for each non-pipelined execution round. The corresponding layer latency is

$$\mathcal{D}_{\text{layer}} = \text{cycle}_{\text{unpipe}}\, k\, \tau, \tag{24}$$

where $\text{cycle}_{\text{unpipe}}$ denotes the number of execution rounds required by the non-pipelined schedule.

When $n_{\text{onchip}} \geq n_{\text{memcover}}$, the off-chip memory cannot simultaneously supply data for all on-chip neurons. The neurons must therefore be divided into multiple memory-covered groups. The number of loading intervals required to cover the available on-chip neurons is

$$\text{incycle}_{\text{pipe}} = \left\lceil \frac{n_{\text{onchip}}}{n_{\text{memcover}}} \right\rceil. \tag{25}$$

Instead of waiting until all groups are loaded before starting the next stochastic bit, the proposed strategy staggers the computations of different bit positions. As indicated by the overlapped loading and computing regions in Fig. 1(b) and Fig. 1(c), while one neuron group is being supplied with new data from DDR, previously loaded groups can continue processing other stochastic bits. This inter-bit overlap converts otherwise idle memory-waiting intervals into useful computation.

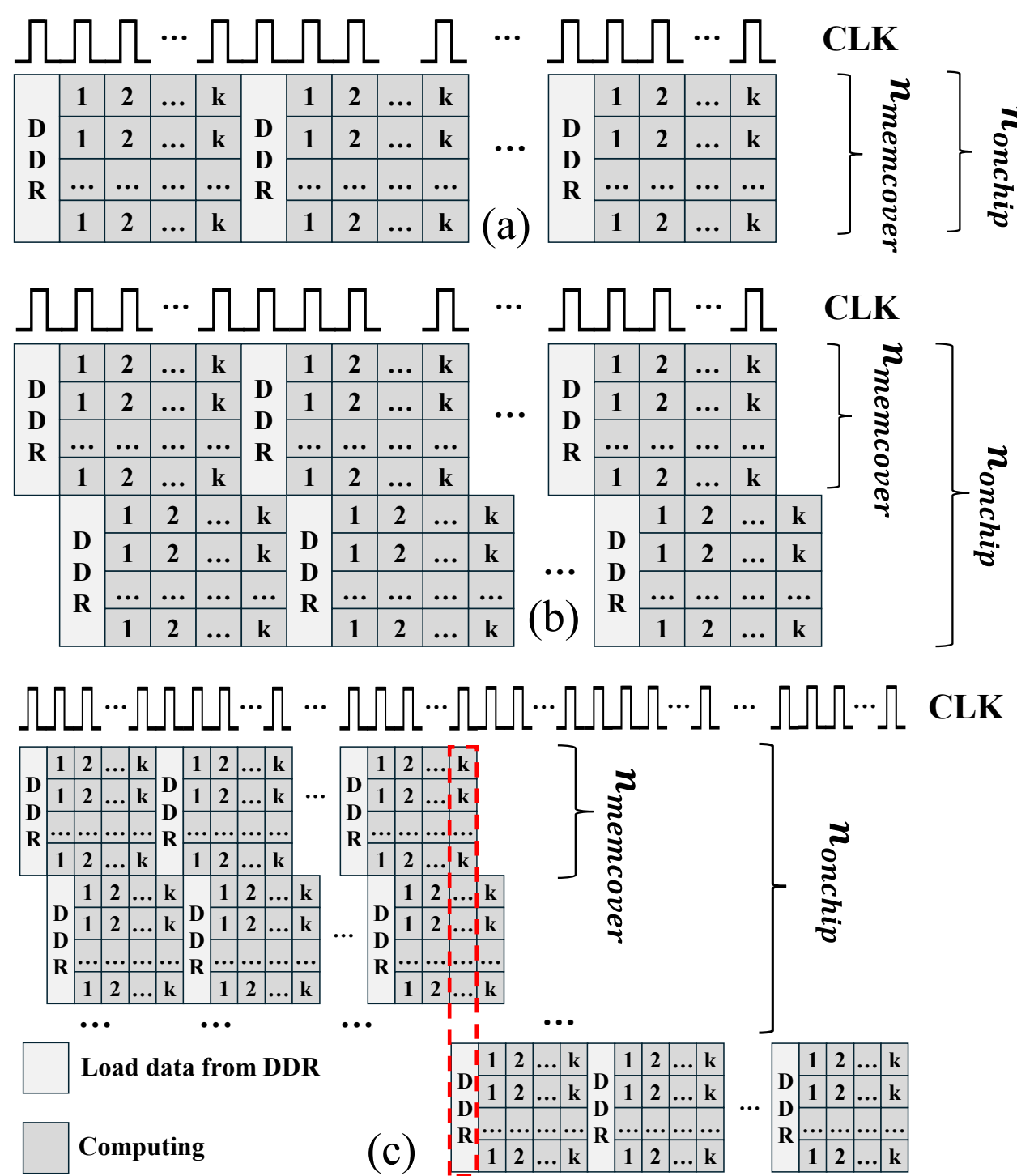

Fig. 1. Proposed pipeline strategy under different application scenarios: (a) no pipeline, (b) partial pipeline, and (c) full pipeline.

If $\text{incycle}_{\text{pipe}} < k$, the required group-filling interval is shorter than the stochastic bitstream length, resulting in the partial-pipeline case shown in Fig. 1(b). In this regime, the pipeline can be filled before all $k$ bit positions have been processed. Consequently, most memory-loading intervals are hidden by computation after the initial filling phase, and the layer latency is

$$\mathcal{D}_{\text{layer}} = \left[\text{cycle}_{\text{pipe}}(k+1) + \text{incycle}_{\text{pipe}} - 1\right]\tau, \quad (26)$$

where $\text{cycle}_{\text{pipe}}$ denotes the number of execution rounds under the pipelined schedule. The term $\text{incycle}_{\text{pipe}} - 1$ accounts for the additional pipeline filling overhead before a steady overlap between memory loading and computation is established.

In contrast, when $\text{incycle}_{\text{pipe}} \geq k$, the number of memory-covered groups is comparable to or larger than the available $k$ stochastic bit positions. This corresponds to the full-pipeline case in Fig. 1(c). In this situation, all bit positions are already involved in the interleaved schedule before the memory interface completes the coverage of all required on-chip neuron groups. Therefore, increasing the overlap across stochastic bits can no longer completely hide the data-loading latency, and the execution becomes strongly constrained by the off-chip

**Algorithm 1 Pipeline Strategy**

**Input:** A layer $\mathcal{L}$ of convolutional neural network.
**Output:** A pipeline strategy for resource allocation and data processing.

1: $n_{onchip}$: available neurons on chip for current layer
2: $n_{memcover}$: neurons covered under current memory load capability
3: $\mathcal{D}_{layer}$: current layer's delay
4: $\tau$: system clock period
5: $k$: stochastic bitstream length
6: **if** $n_{onchip} < n_{memcover}$ **then**
7:     no pipeline, $n_{used} = n_{onchip}$
8:     $\mathcal{D}_{layer} = cycle_{unpipe} \times k \times \tau$
9: **else**
10:     pipeline can be used, $n_{used} < n_{onchip}$
11:     $incycle_{pipe} = ceil(n_{onchip}/n_{memcover})$
12:     **if** $incycle_{pipe} < k$ **then**
13:         partially pipelined
14:         $\mathcal{D}_{layer} = [cycle_{pipe} \times (k+1) + incycle_{pipe} - 1] \times \tau$
15:     **else**
16:         fully pipelined
17:         $\mathcal{D}_{layer} = (k + incycle_{pipe}) \times \tau$
18:     **end if**
19: **end if**

bandwidth. According to Algorithm 1, the resulting layer latency is

$$\mathcal{D}_{\text{layer}} = \left(k + \text{incycle}_{\text{pipe}}\right)\tau. \quad (27)$$

Overall, the three cases adapt the execution schedule to the relationship among on-chip parallelism, off-chip memory coverage, and stochastic bitstream length, thereby improving logic-resource utilization without requiring an increase in external memory bandwidth.